\documentclass[twocolumn,superscriptaddress,amssymb,amsmath,nobibnotes,aps,prd,showpacs,nofootinbib]{revtex4}%
\pdfoutput=1

\usepackage{graphicx,float}
\usepackage{epsf}
\usepackage{bm}
\usepackage{amsmath,hyperref}
\usepackage{amsfonts}
\usepackage{amssymb}
\usepackage{color}
\definecolor{darkgreen}{rgb}{0., 0.65, 0.1}
\begin{document}

\title{Dark sector interaction: A remedy of the tensions between CMB and LSS data}

\author{Suresh Kumar}
\email{suresh.kumar@pilani.bits-pilani.ac.in}
\affiliation{Department of Mathematics, BITS Pilani, Pilani Campus, Rajasthan-333031, India}
 
\author{Rafael C. Nunes}
\email{rafadcnunes@gmail.com}
\affiliation{Divis\~ao de Astrof\'isica, Instituto Nacional de Pesquisas Espaciais, Avenida dos Astronautas 1758, S\~ao Jos\'e dos Campos, 12227-010, SP, Brazil}

\author{Santosh Kumar Yadav}
\email{sky91bbaulko@gmail.com}
\affiliation{Department of Mathematics, BITS Pilani, Pilani Campus, Rajasthan-333031, India}


\begin{abstract}
The well-known tensions on the cosmological parameters $H_0$ and $\sigma_8$ within the $\Lambda$CDM cosmology shown by the Planck-CMB and LSS data are possibly due to the systematics in the data or our ignorance of some new physics beyond the $\Lambda$CDM model. In this letter, we focus on the second possibility, and investigate a minimal extension of the $\Lambda$CDM model by allowing a  coupling between its dark sector components (dark energy and dark matter). We analyze this scenario with Planck-CMB, KiDS and HST data, and find that the $H_0$ and $\sigma_8$ tensions disappear. In the joint analyses with Planck, HST and KiDS data, we find non-zero coupling in the dark sector up to 99\% CL. Thus, we find a strong statistical support from the observational data for an interaction in the dark sector of the Universe while solving the $H_0$ and $\sigma_8$ tensions simultaneously.

\end{abstract}

\maketitle

\section{INTRODUCTION}
The cosmic microwave background (CMB) observations from Planck \cite{Planck2015:A13} together with the observations of cosmic expansion history from independent measurements, including baryonic acoustic oscillations (BAO) \cite{BAO} and Supernovae type Ia \cite{SNIe}, find a very good statistical fit to the standard model of cosmology, viz., the $\Lambda$CDM (cosmological constant $\Lambda$ + cold dark matter) model. However, with the gradual increase in the data accumulation with precision, the latest Planck-CMB data show inconsistency with the direct measurements of Hubble expansion rate from the Hubble Space Telescope (HST) \cite{riess}, and some large scale structure (LSS) observations such  as galaxy cluster counts \cite{LSS1,LSS2} and  weak lensing \cite{kids01,kids02}, in the framework of the $\Lambda$CDM model. Specifically, the value of present Hubble constant $H_0$ and the value of r.m.s. fluctuation of density perturbation at $8 h^{-1}$ Mpc scale (characterized by $\sigma_8$), inferred from the CMB experiments are in a serious disagreement with  the ones measured from the LSS experiments \cite{DES01,DES02}. \\

At present, it is not clear whether these inconsistencies are due to systematics in the data or need some physics beyond the standard $\Lambda$CDM model \cite{issue01,issue02}. 
Several studies have been carried out in the literature to reconcile these tensions between the CMB and LSS observations \cite{H01, H02, H03, H04, H05}. But both the tensions are not resolved simultaneously at a significant statistical level. Rather, by assuming neutrinos properties, the parameters are correlated in such  a way that lower values of $\sigma_8$ require higher values of total matter density and smaller values of $H_0$, which aggravates the tensions (e.g. \cite{prl01}). In \cite{prl02}, it is argued that the presence of sterile neutrinos do not bring a new concordance, but possibly indicating systematic biases in the measurements. However, recently in \cite{mohanty01}, it has been argued that incorporation of the dissipative effects in the energy momentum  tensor  can  ameliorate both the tensions simultaneously. Likewise in \cite{mohanty02}, it is claimed that the presence of viscosity, shear or bulk or combination of both, can alleviate both the tensions simultaneously.\\

At present, the precise nature of constituents of the dark sector in the $\Lambda$CDM model, namely CDM and dark energy (the vacuum energy mimicked by $\Lambda$), is unknown. Moreover, these two are major energy ingradients accounting for around 95\% energy budget of the Universe. So, a possibility of exchange of energy/momentum or interaction between the dark sector components can not be ignored, especially considering the current issues with the $\Lambda$CDM model. Consequently, in recent years, a large number of studies have been carried out with regard to the interaction between the dark sector components of the Universe with different motivations and perspectives \cite{IDE01, IDE02, IDE03, IDE04,IDE05,IDE06,IDE07,IDE08,IDE09,IDE10,IDE10a,IDE11,IDE11a,IDE12,IDE13,IDE14,IDE15,IDE16} (see \cite{review02} for a review). In particular, a possible interaction in the dark sector has been investigated in \cite{IDE02,IDE03,IDE04,IDE05}, where it has been  argued that a  dark sector coupling could be a possible remedy to the $H_0$ and $\sigma_8$ tensions.\\

In this letter, we investigate a minimal extension of the $\Lambda$CDM model by allowing interaction among its dark sector ingredients with the Planck-CMB, KiDS and HST data. The main result of this work is that 
the $H_0$ and $\sigma_8$ tensions that prevail within the $\Lambda$CDM model, disappear 
in presence of a dark sector coupling, while we find non-zero coupling between dark matter and dark energy up to 99\% CL. Thus, we find a strong statistical support from the observational data for an interaction in the dark sector of the Universe while alleviating the $H_0$ and $\sigma_8$ tensions simultaneously. In what follows, we present details of the model, data sets with the methodology of analysis, results with the discussion, and some final remarks.

\section{Interaction in the dark sector} In general, the background evolution of coupled dark sector components, in the Friedmann-Lema\^{i}tre-Robertson-Walker Universe,  is encoded in the coupled energy-momentum conservation equations:
\begin{equation}
\label{Tuv}
\dot{\rho}_{\rm dm} + 3 H \rho_{\rm dm} = - \dot{\rho}_{\rm de} - 3 H \rho_{\rm de} (1 + w_{\rm de}) = Q,
\end{equation}
where an over dot stands for the cosmic time derivative;  $\rho_{\rm dm}$ and $\rho_{\rm de}$ are the energy densities of dark matter and dark energy, respectively; $H =\dot{a}/a$ is the Hubble parameter with $a$ being
the scale factor of the Universe; $w_{\rm de}$ is equation of state parameter of dark energy; and $Q$ is the coupling function between the dark sector components, which characterizes the interaction form, viz., $Q < 0$ corresponds to energy flow from dark matter to dark energy, and $Q > 0$ the opposite case.
The most commonly used forms of $Q$ in the literature are: $Q \propto H \rho_{\rm dm}$ or $Q \propto H \rho_{\rm de}$ or their combinations  \cite{Wang08, Hai-Li}. In this letter, we use $Q \propto H \rho_{\rm de}$  in order to avoid the  instability in the perturbations at early times \cite{Wang09,Valiviita09}. Thus, we use the form $Q =\delta H \rho_{\rm de}$, where $\delta$ is the coupling parameter that quantifies the coupling between dark matter and dark energy.

At perturbative level,  we adopt the synchronous gauge in which the evolution  of  the  scalar  mode  perturbations  within a general interacting dark matter and dark energy scenario, in the Fourier space, is governed by the equations \cite{Wands01, Wands02,wands19}:

\begin{equation}
\label{delta}
\dot{\delta}_{\rm dm} - \frac{k^2}{a^2} \theta_{\rm dm} + \frac{\dot{h}}{2} - \frac{Q}{\rho_{\rm dm}} \delta_{\rm dm} = \frac{\dot{\delta}_{\rm de}}{\rho_{\rm dm}},
\end{equation}

\begin{equation}
\label{theta}
\dot{\theta}_{\rm dm} \rho_{\rm dm} =  \delta_{\rm de} + Q \theta_{\rm dm}.
\end{equation}
where $Q$ is the previously defined coupling function and $h$ is the scalar mode in synchronous gauge. In addition, we assume the energy transfer flow between the dark sector components parallel to the four-velocity of the dark matter, i.e., $Q^{\mu}_{\rm dm} = - Q u^{\mu}_{\rm dm}$. Thus, there is no momentum transfer in the rest frame of dark matter, and the velocity perturbation for dark matter is not affected by the interaction, and therefore obeys the standard evolution as expected in the synchronous gauge. Therefore, the dark matter four-velocity $u^{\mu}_{\rm dm}$ is a geodesic flow, i.e, $u^{\mu}_{\rm dm} \nabla_{\mu} u^{\nu}_{\rm dm} = 0$. A direct consequence is that the vacuum energy perturbation contribution in the dark matter-comoving frame is identically null. The other species (baryons, photons and neutrinos) are conserved independently, and their dynamics follow the well-known standard evolution both at the background and perturbative levels.

Following the above arguments, in this work, we adopt $w_{\rm de} = - 1$, i.e., we allow the interaction of vacuum energy with the CDM, and refer this scenario to as IVCDM model in the remaining text. This model is investigated in  many studies eg. \cite{IDE03,Wands01, Wands02}, and very recently in \cite{wands19}, but mainly in the context of interaction in the dark sector\footnote{The preprint \cite{wands19} appeared on arXiv during the final stage of our study.}. Here, we present an analysis with the main objective to investigate whether this said dark sector interaction could be a possible remedy of the tensions between the CMB and LSS data.\

\section{Data sets and methodology of the analysis} To analyze the IVCDM model in contrast with the $\Lambda$CDM model, we use the following observational data sets: (i) Planck: CMB temperature and polarization data from Planck-2015 \cite{Planck2015:A13} comprised of likelihoods of low-$\ell$
temperature and polarization likelihood at $\ell\leq 29$,
temperature (TT) at $\ell\geq 30$, (ii) KiDS: the  measurements of the weak gravitational lensing shear power spectrum from the Kilo Degree Survey \cite{kids01}, and (iii) HST: the new local value of Hubble constant $H_0 =73.24 \pm 1.74$  km s${}^{-1}$ Mpc${}^{-1}$ \cite{riess}.
We use the publicly available Boltzmann code \texttt{CLASS} \cite{class} with the parameter inference code \texttt{Monte python} \cite{monte} to obtain correlated Monte Carlo Markov Chain (MCMC) samples. We follow the Gelman-Rubin  convergence criterion \cite{Gelman_Rubin} of MCMC chains, requiring $1-R<0.03$, for all the model parameters. We use the MCMC samples analysing python package \texttt{GetDist} \cite{antonygetdist}. In all analyses performed here, we choose uniform priors on $\Lambda$CDM and IVCDM baseline parameters: $\omega_{\rm b}$, $\omega_{\rm cdm}$,  $A_{s}$, $n_s$, $H_0$ and $\delta$, as shown in Table \ref{tab_results}. We analyze the two models with Planck and KiDS data separately to clearly demonstrate the issue/resolution of the tensions among the two data sets. In order to obtain more tight constraints on the model parameters, we also study two joint analyses with HST data: Planck + HST and Planck + HST + KiDS.

\section{Results and discussion}
Table \ref{tab_results} summarizes the main results from the statistical analyses of the $\Lambda$CDM and IVCDM models with Planck, KiDS, Planck + HST and Planck + HST + KiDS data. We notice similar constraints on the baseline parameters $\omega_{\rm b}$, $\omega_{\rm cdm}$, $A_{s}$, $n_s$, in the two models in all the four cases of data sets under consideration. In what follows, we discuss the constraints on other parameters with regard to the tensions on the parameters $H_0$ and $\sigma_8$, in particular. \\

\begin{table*}[htb!] 
\caption{\label{tab_results} {Constraints (68\% CL)  on  free and some derived parameters of the $\Lambda$CDM and IVCDM models from the four data combinations.  The final row displays $\chi^2_{\rm min}$ values of the statistical fit.}}
\begin{tabular} {c| c| c c | c c| c c | c c }  
\hline \hline  

 Parameter& Prior & \multicolumn{2}{c|}{Planck} & \multicolumn{2}{c|}{KiDS} & \multicolumn{2}{c|}{Planck + HST} & \multicolumn{2}{c}{ Planck + HST + KiDS}  \\  [0.7ex]
 \hline
 &  &  $\Lambda$CDM   & IVCDM  &    $\Lambda$CDM   &  IVCDM  &  $\Lambda$CDM   &  IVCDM  &    $\Lambda$CDM   & IVCDM   \\  [0.7ex]
 \hline
 $10^{2}\omega_{\rm b }$ &$[1.9,2.6]$ &  $2.23^{+ 0.02}_{-0.02}$  &  $2.22^{+ 0.02}_{-0.02}$  &   $2.23^{+0.20}_{-0.20}  $&   $2.25^{+0.20}_{-0.20} $ 
 
  &  $2.25^{+ 0.02}_{-0.02}$  &  $2.22^{+ 0.02}_{-0.02}$  &   $2.26^{+0.20}_{-0.20}  $  &   $2.23^{+0.02}_{-0.02}  $     \\[1ex]
 
 $\omega_{\rm cdm }  $ &$[0.01,0.99]$ &    $0.120^{+0.002}_{-0.002}   $  &    $0.120^{+0.002}_{-0.002}   $  &  $0.124^{+0.040}_{-0.046}$ &  $0.123^{+0.042}_{-0.042}$ 
 
 &    $0.120^{+0.002}_{-0.002}   $ &    $0.120^{+0.002}_{-0.002}   $  &  $0.115^{+0.001}_{-0.001}$  &  $0.119^{+0.002}_{-0.002}$ \\[1ex]

$\ln[10^{10}A_{s }]$    &$[1.7,5]$ &  $3.120^{+0.006}_{-0.006}  $    &  $3.121^{+0.007}_{-0.007}  $  & $2.760^{+0.510}_{-1.000}   $   & $2.800^{+0.400}_{-1.100}  $  

&  $3.116^{+0.006}_{-0.006}  $   &  $3.120^{+0.006}_{-0.006}  $     & $3.114^{+0.006}_{-0.016}   $ & $3.120^{+0.006}_{-0.006}   $  \\[1ex] 

$n_{s } $ & $[0.7,1.3]$ &   $0.967^{+0.005}_{-0.005}  $  &    $0.965^{+0.005}_{-0.005}  $  & $1.060^{+0.220}_{-0.098}$    & $1.070^{+0.210}_{-0.092}$ & 

$0.973^{+0.005}_{-0.005}  $    &    $0.964^{+0.006}_{-0.006}  $  & $0.978^{+0.005}_{-0.005}$  & $0.967^{+0.006}_{-0.006}$  \\[1ex]

$  H_0 $  & $[60,90]$&  $ 67.8^{+0.9}_{-0.9}$   &  $72.2^{+3.5}_{-5.0}$    &   $73.6^{+7.8}_{-3.6}$       &   $74.2^{+7.5}_{-5.1}	$  &

$68.9^{+0.8}_{-0.8}$   &  $72.9^{+1.7}_{-1.7}$         &   $69.7^{+0.7}_{-0.7}$   &   $73.6^{+1.6}_{-1.6}$  \\[1ex]

 $\delta $   &$[-1, 1]$ & $0$      &  $ -0.34^{+0.40}_{-0.26}$  &  $0$   & $-0.23^{+0.43}_{-0.43}$   
 
 &  $0$     &  $ -0.40^{+0.17}_{-0.14}$ &  $0$     & $-0.40^{+0.16}_{-0.14}$   \\ [1ex]
\hline
$\Omega_{\rm{m0} }$  &-- & $0.309^{+ 0.012}_{-0.012} $    &  $0.276^{+ 0.031}_{-0.031} $    & $0.274^{+0.074}_{-0.094}  $ & $0.267^{+0.072}_{-0.094}	  $  

&  $0.294^{+ 0.010}_{-0.010} $   &  $0.269^{+ 0.012}_{-0.014} $  & $0.284^{+0.008}_{-0.008}  $  & $0.262^{+0.010}_{-0.012}  $  \\[1ex]
  
 $\sigma_{8} $          &-- & $0.838^{+ 0.007}_{-0.007}  $  & $0.725^{+ 0.140}_{-0.072}  $    & $ 0.734^{+ 0.086}_{-0.170}  $  & $0.678^{+0.080}_{-0.230}	 $ 
 
 & $0.830^{+ 0.007}_{-0.007}  $  & $0.710^{+ 0.054}_{-0.045}  $ & $ 0.824^{+ 0.007}_{-0.006}  $ & $ 0.702^{+ 0.049}_{-0.049}  $ \\[1ex]
 
 
 \hline
 $\chi^2_{\rm min}/2   $  &-- & 5631.59   &   5631.75      & 24.06       & 24.21       
 & 5635.52  & 5631.69  & 5662.63  & 5659.82   \\[1ex]  
 
\hline \hline
  \end{tabular}
\end{table*}

\begin{figure}[h]
\includegraphics[width=4.2cm]{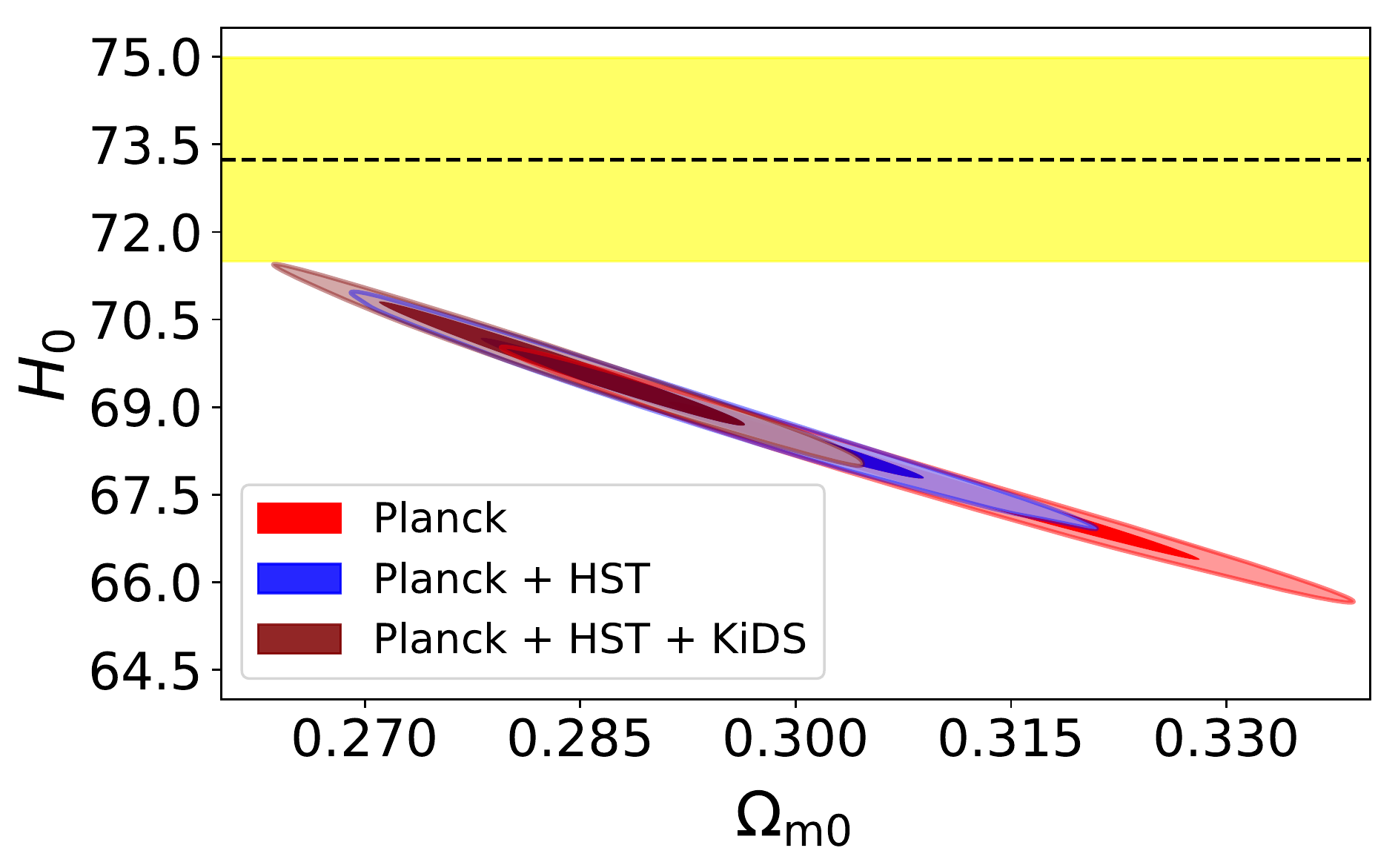}
\includegraphics[width=4.2cm]{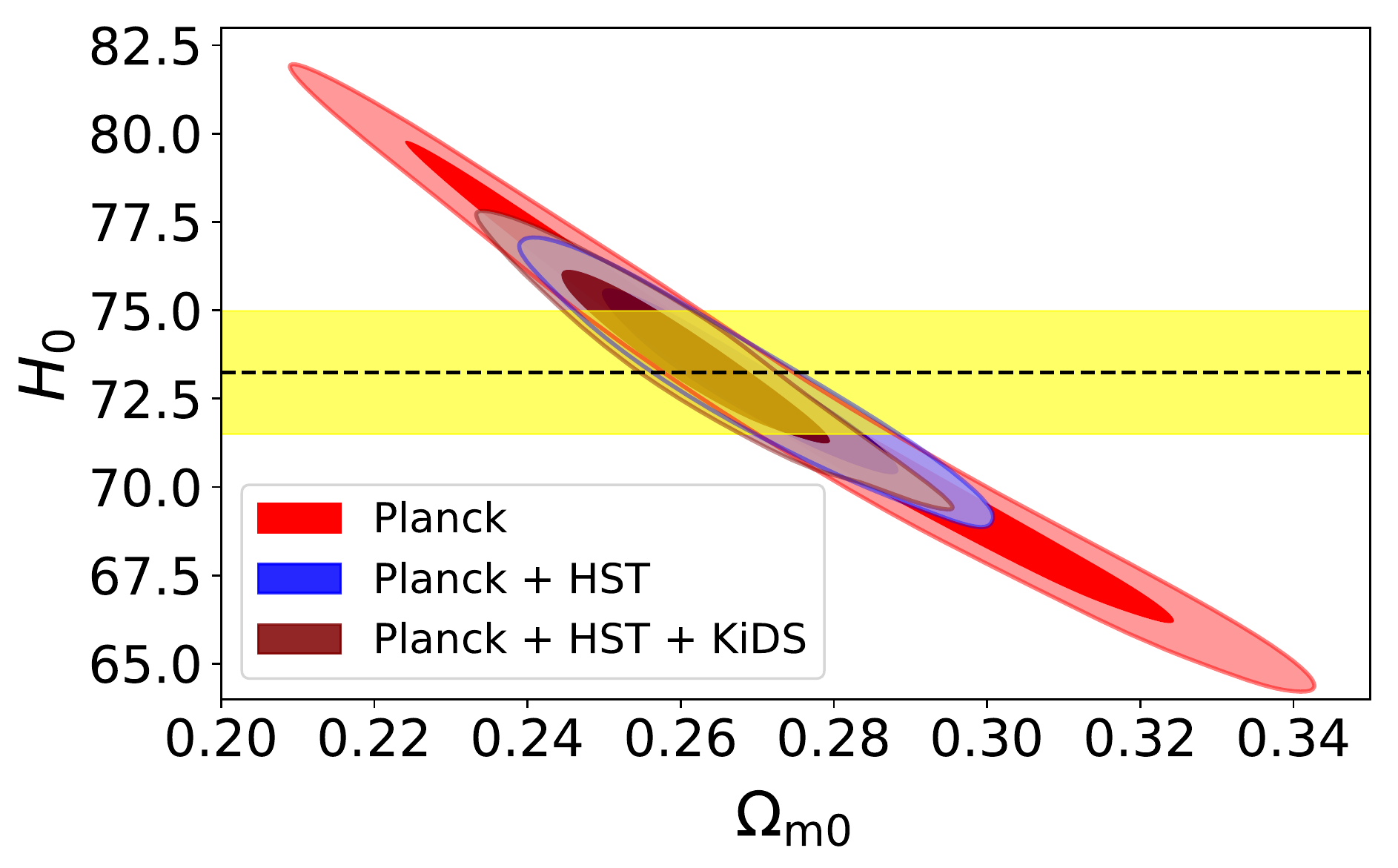}
\caption{Parametric space (68\% and 95\% CL) in the plane $\Omega_{\rm m0}-H_0$ for the $\Lambda$CDM (left panel) and IVCDM model (right panel) from three data sets. In the left panel, it is clear to see that the local measurement of $H_0 =73.24 \pm 1.74$ km s$^{-1}$Mpc$^{-1}$ \cite{riess} (yellow band) is in disagreement with the statistical region of $H_0$ from the three data sets within the $\Lambda$CDM cosmology. In the right panel, we see that there is no tension on $H_0$ within 68\% CL in the IVCDM model.}\label{h0tension}
 \end{figure}

First we discuss the constraints with regard to the tension on $H_0$. In the left panel of Fig. \ref{h0tension}, the $\Omega_{\rm m0}-H_0$ parametric space is shown for the $\Lambda$CDM model with a yellow band corresponding to the local value $H_0 =73.24 \pm 1.74$ km s$^{-1}$ Mpc$^{-1}$ \cite{riess}, in case of Planck, Planck + HST and Planck + HST + KiDS data\footnote{We have not shown the $\Omega_{\rm m0}-H_0$ statistical region for KiDS data set because it is insensitive to the parameter $H_0$ \cite{kids01}.}. Clearly, the local measurement of $H_0$ is in disagreement with the region of $H_0$ predicted by Planck data \cite{Planck2015:A13}, and other two data combinations within the $\Lambda$CDM cosmology. In the right panel of Fig. \ref{h0tension}, the $\Omega_{\rm m0}-H_0$ parametric space is shown for the IVCDM model, where we observe that there is no tension on $H_0$.  


\begin{figure}[h]
\includegraphics[width=4.2cm]{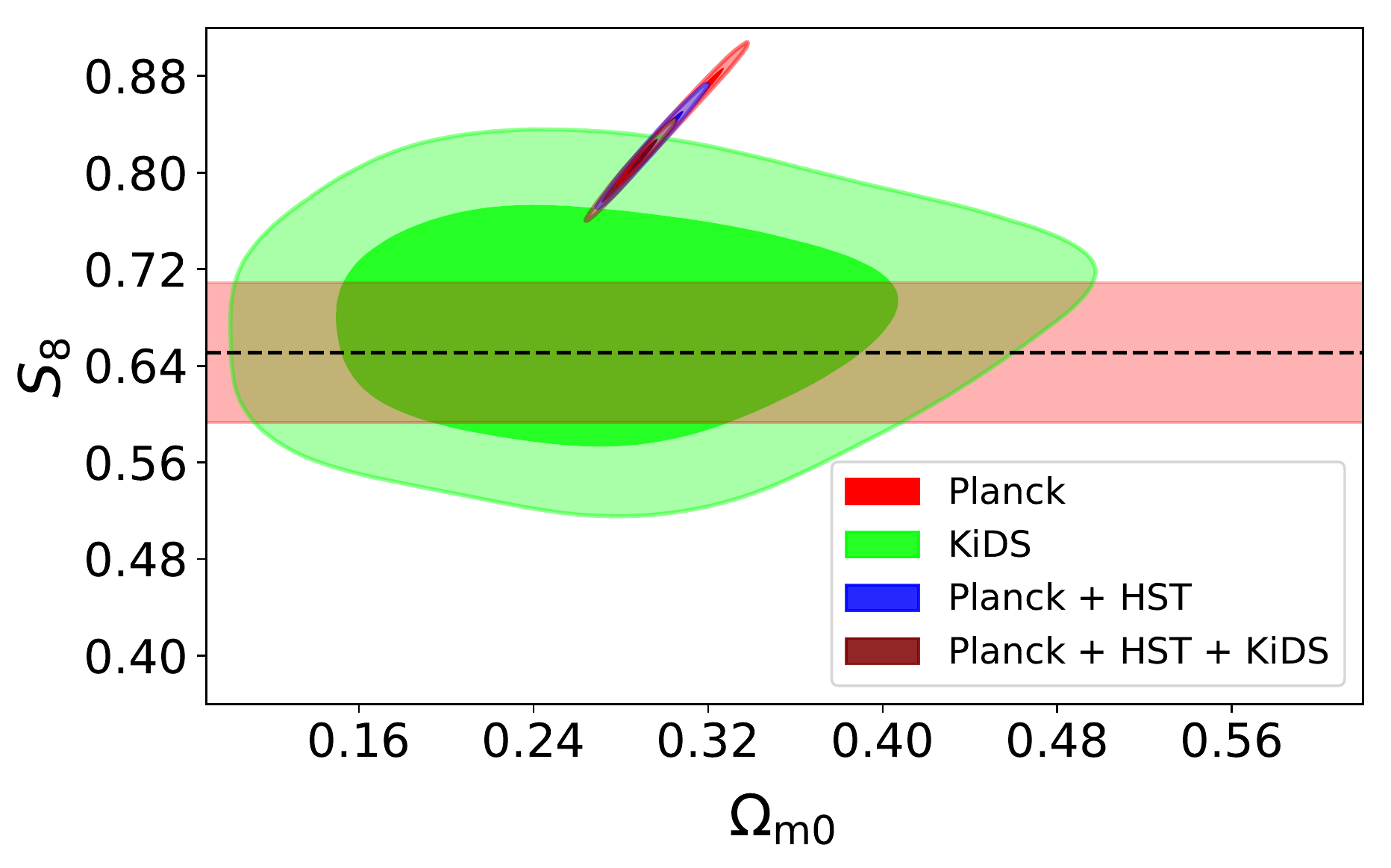}
\includegraphics[width=4.2cm]{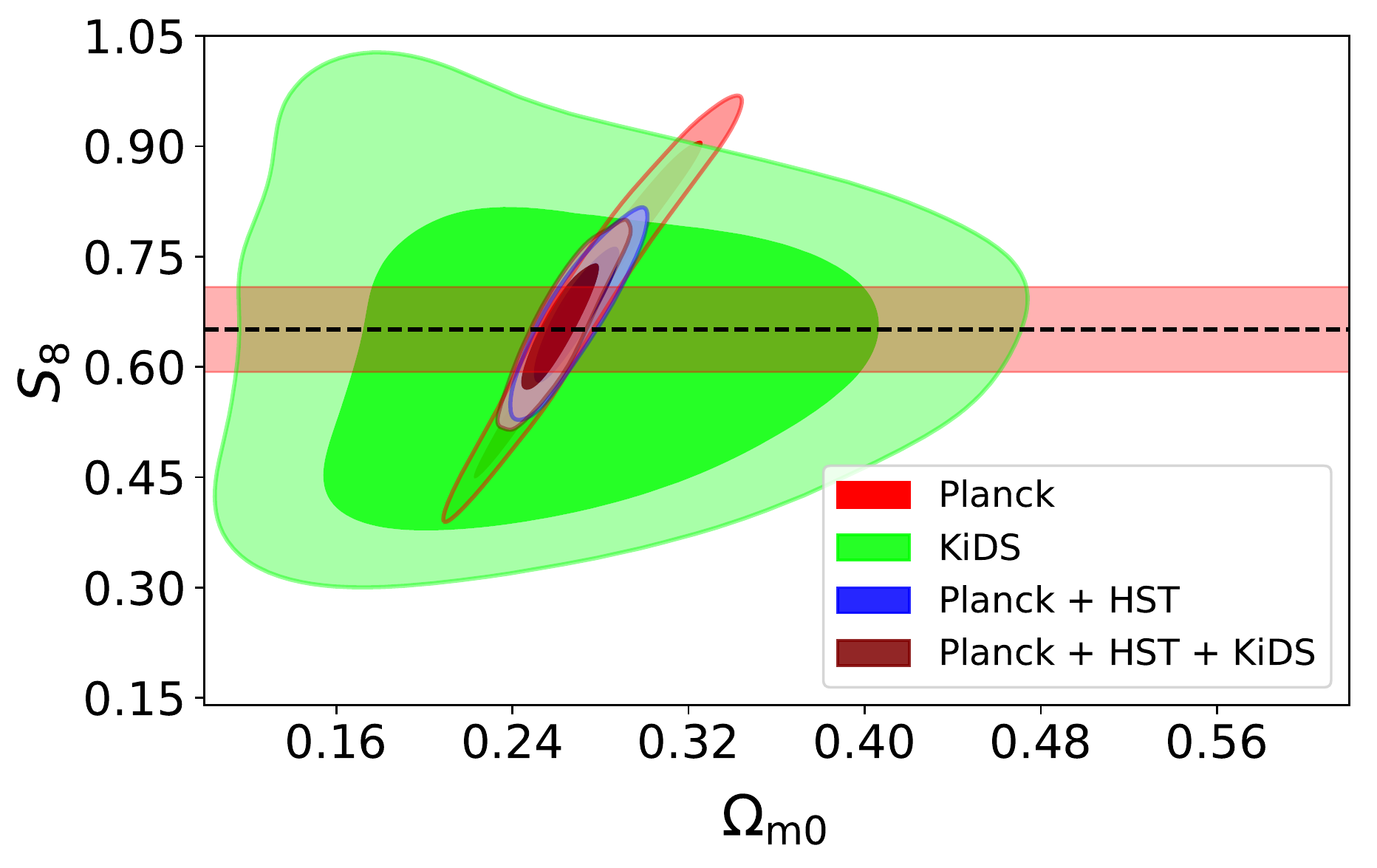}
\caption{Parametric space (at 68\% and 95\% CL) in the plane $\Omega_{\rm m0}-S_8$ for the $\Lambda$CDM (left panel) and IVCDM model (right panel) from four data sets. In the left panel, we see that the $\Omega_{\rm m0}-S_8$ region given by the Planck data within the $\Lambda$CDM cosmology is clearly in disagreement with the region $S_8 \equiv \sigma_8 \sqrt{\Omega_{\rm m0}/{0.30}} = 0.651 \pm 0.058$, predicted by KiDS data \cite{kids01} (red band). In the right panel for the IVCDM model, we observe that there is no tension on $S_8$ within 68\% CL.}
\label{S8tension}
\end{figure}


With regard to the tension on $\sigma_8$, in the left panel of Fig. \ref{S8tension}, the $\Omega_{\rm m0}-S_8$ parametric space is shown for the $\Lambda$CDM model from the four analyses performed here. Clearly, region given by Planck and Planck + HST data within the $\Lambda$CDM cosmology is in disagreement with the region predicted by KiDS data. In the right panel of Fig. \ref{S8tension}, we show the the same parametric space for the IVCDM model, where we note that there is no tension on $S_8$, and all these data sets are in agreement with each other. It is important to note that, since the CMB and LSS predictions are not in tension with each other within the IVCDM model, we can use all these data in a joint analysis. 

\begin{figure}[hbt!]
\includegraphics[width=4.2cm]{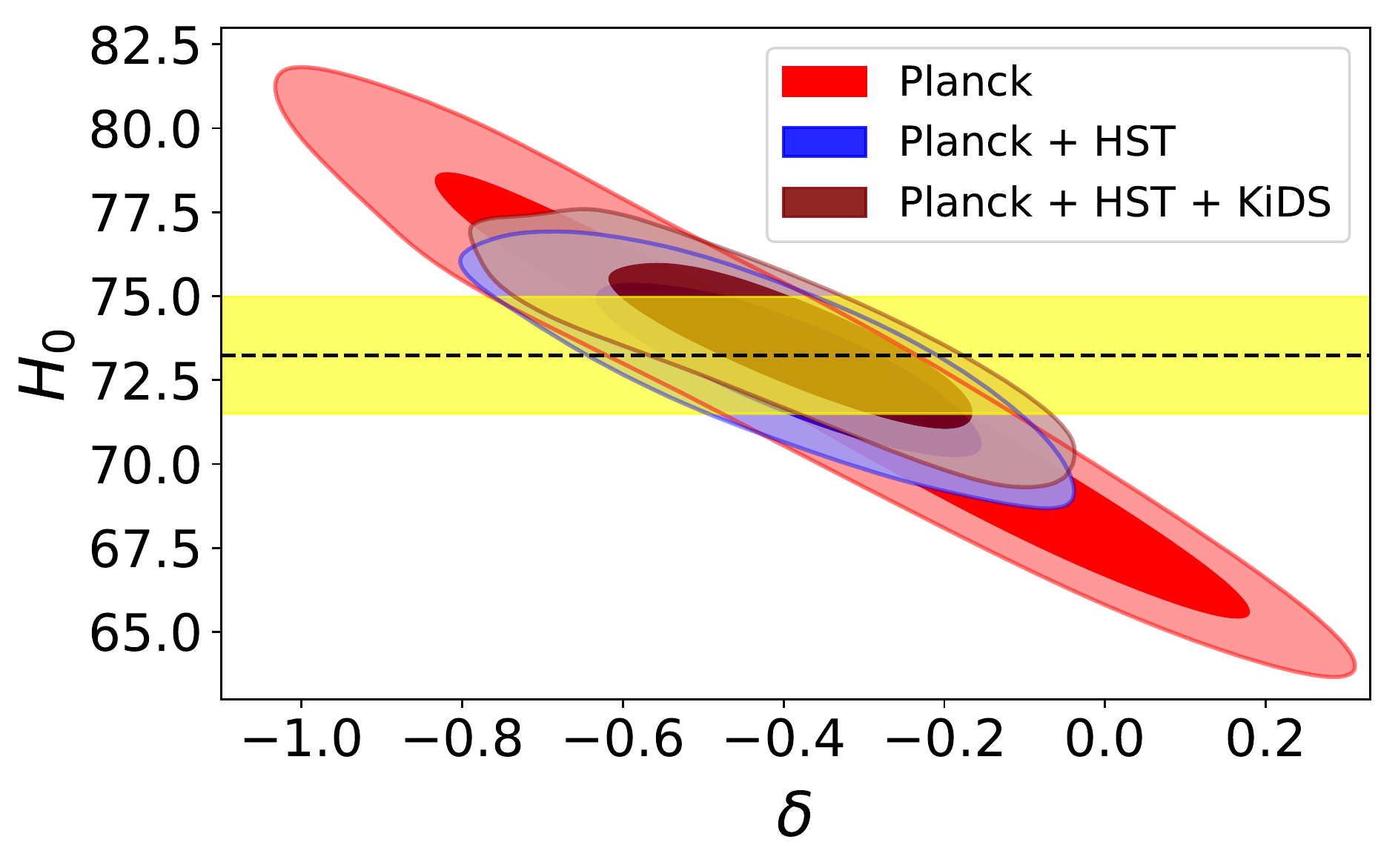}
\includegraphics[width=4.2cm]{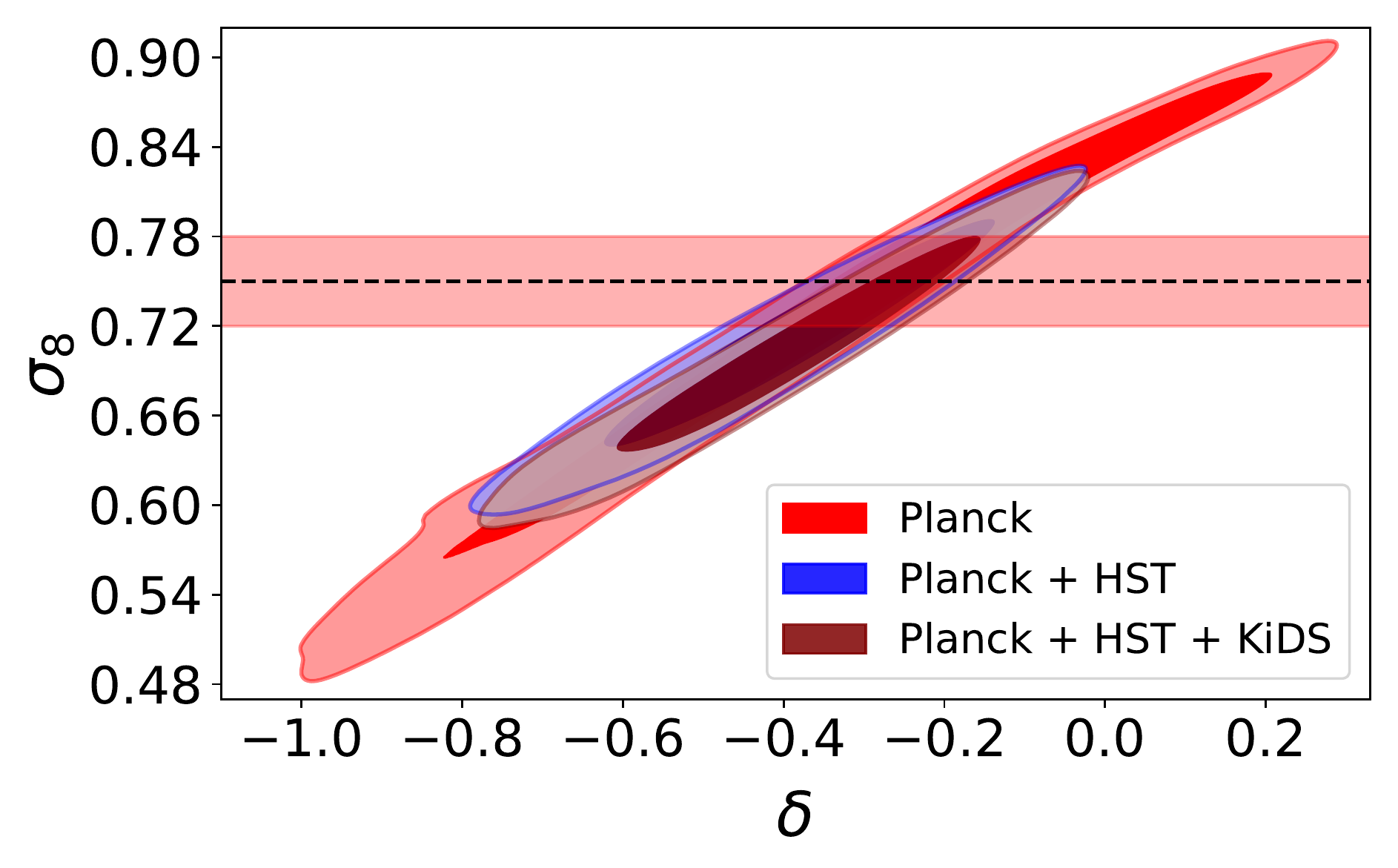}
\caption{$\delta-H_0$ (left panel) and $\delta-\sigma_8$ (right panel) parametric spaces (68\% and 95\% CL) for the IVCDM model from three data sets. The yellow band corresponds to local value $H_0 =73.24 \pm 1.74$ km s$^{-1}$Mpc$^{-1}$ \cite{riess} whereas the light red band corresponds to $\sigma_8 = 0.75 \pm 0.03$ \cite{LSS2}.}\label{hs8_delta}
\end{figure}

In the above, we have shown that the tensions on both the parameters $H_0$ and $\sigma_8$ disappear within the framework of IVCDM model. Next, we focus our attention on the coupling parameter $\delta$. In Fig. \ref{hs8_delta}, the statistical regions (at 68\% and 95\% CL) on $\delta$ are shown with $H_0$ and $\sigma_8$ from Planck data, and the other two data sets including Planck. We observe that $\delta$ finds a negative correlation with $H_0$ while a positive correlation with $\sigma_8$. It amounts to saying that lower values of $\delta$ correspond to higher values of $H_0$ and lower values of $\sigma_8$, which is nice with regard to resolving tensions on the both $H_0$ and $\sigma_8$ simultaneously.  The correlation strength $r$ of $\delta$ with $H_0$ and $\sigma_8$ is quantified in Table \ref{correlation} in case of all the four data sets. We notice very strong correlations of $\delta$ with $H_0$ and $\sigma_8$ in case of Planck data, and two other data combinations with the Planck data. Interestingly, $\delta$ shows a strong and positive correlation with $\sigma_8$ in case of KiDS data, as well. 

\begin{table}[h]
    \centering
        \caption{Correlation $r$ of $\delta$ with $H_0$ and $\sigma_8$.}\label{correlation}
    \begin{tabular}{c c c}
    \hline \hline
    Data& $r_{\delta H_0}$ & $r_{\delta\sigma_8}$  \\
    \hline
     Planck  &  $-0.9662$   &  0.9810 \\
     KiDS &   $\;\;\;0.0397$  &  0.7768 \\
     Planck + HST  &  $-0.8205$ &  0.9595 \\
     Planck + HST + KiDS   & $-0.8463$ &  0.9672  \\
    \hline \hline
\end{tabular}

\end{table}

We find at 99\% CL on $\delta$, viz., $-0.34^{+0.59}_{-0.65}$, $-0.23^{+0.72}_{-0.77}$, $-0.40^{+0.35}_{-0.44}$, and $-0.40^{+0.36}_{-0.41}$  for the Planck, KiDS, Planck + HST, Planck + HST + KiDS data, respectively.  We notice that the mean values of $\delta$ in all cases are negative, indicating the energy/momentum flow from the dark matter to dark energy. Clearly, it is reflected by the lower values of $\Omega_{\rm m0}$ in the IVCDM model compared to the $\Lambda$CDM model in all cases displayed in Table \ref{tab_results}. Further, it is interesting to observe that the non-null range of $\delta$ with negative values is up to $99\%$ CL in the joint analyses. Thus, we find a strong statistical support from the data for interaction in the dark sector of the $\Lambda$CDM Universe while alleviating the $H_0$ and $\sigma_8$ tensions simultaneously. These results are interesting, and the model is well-behaved both at background and perturbative levels.

Finally, we  perform a statistical comparison of the IVCDM model with the $\Lambda$CDM model by using the well-known Akaike Information Criterion (AIC) \cite{AIC01, AIC02}: 
  \begin{equation}
  \text{AIC} = -2 \ln  \mathcal{L}_{\rm max} + 2 N \quad = \chi_{\rm min}^2 + 2N,
 \end{equation}
where $ \mathcal{L}_{\rm max}$ is the maximum likelihood function of the model, and $N$ is the total number of free parameters in the model. For the statistical comparison, the AIC difference between the model under study and  the reference model is calculated. This difference in AIC values can be interpreted as the evidence in favor of the model under study over the reference model. It has been argued in \cite{AIC03} that one model can be considered as better  with respect to other if the AIC difference between the  two models is greater than a threshold value $\Delta_{\rm threshold}$. According to thumb rule of AIC, $\Delta_{\rm threshold} = 5$ is the universal threshold value (the minimum AIC difference value \cite{Liddle}) to assert a strong support in favor of a model compared to other, regardless the properties of the model considered for comparison. Thus, an AIC difference of 5 or more between two models favors the model with smaller AIC value.
 
  \begin{table}[hbt!]
\caption{\label{evidence}{}Difference of AIC values  of the IVCDM model with respect to reference model ($\Lambda$CDM) with all the data combinations used in this work: $ \Delta \rm AIC =  \rm AIC (IVCDM) -\rm AIC (\Lambda CDM) $. }
\begin{tabular}{c c }
\hline \hline
Data  &  $ \Delta \rm AIC$     \\
\hline
Planck              & $\;\;\;2.32$      \\
KiDS                & $\;\;\;2.30$     \\
Planck + HST        & $-5.66$     \\
Planck + HST + KiDS  & $-3.62$      \\
\hline \hline
\end{tabular}
 \end{table}

Table \ref{evidence} summarizes the AIC differences of IVCDM model with reference model ($\Lambda$CDM) for the four data combinations. One may notice that in all the analyses performed here, we do not find any strong support in favor of the $\Lambda$CDM model. On the other hand, in general, the IVCDM model is penalized in the AIC criterion due to one extra free parameter when compared to the $\Lambda$CDM model. Interestingly, it overcomes the said penalty in case of the Planck + HST data, and finds strong preference over the $\Lambda$CDM model. Also, we observe a mild preference of the IVCDM model in case of the Planck + HST + KiDS data. Thus, the AIC criterion favors the IVCDM model over the $\Lambda$CDM model in the two joint analyses.
 
\section{Final remarks}
In this work, we have demonstrated that a simple and minimal extension of the $\Lambda$CDM model via a coupling between the dark sector ingredients alleviates the tensions on the parameters $H_0$ and $\sigma_8$ simultaneously with excellent accuracy. Also, we have found a possible non-null coupling in the dark sector up to 99\% CL in the joint analyses which amounts to indicating a strong statistical support from the observational data for the dark sector coupling. Therefore, it is clear that a possible interaction 
between dark matter and dark energy is a viable remedy for the tensions in the cosmological data. Indeed, the results presented here are interesting, and therefore general aspects of the dark sector interaction model deserve further investigations.

\begin{center}
\textbf{Acknowledgments}
\end{center}

The authors thank the referee for his/her valuable comments. S.K. gratefully acknowledges the support from SERB-DST project No. EMR/2016/000258, and DST FIST project No.
SR/FST/MSI-090/2013(C). S.K.Y. acknowledges the Council of Scientific \& Industrial
Research (CSIR), Govt. of India, New Delhi, for awarding Senior Research Fellowship (File No. 09/719(0073)/2016-EMR-I).

\end{document}